\documentclass[onecolumn,aps,floatfix,prd,notitlepage,superscriptaddress,showpacs]{revtex4-1}

\usepackage{amsmath}
\usepackage{amssymb}
\usepackage{graphicx,psfrag}
\usepackage{dcolumn}
\usepackage{bm}
\usepackage{xcolor}
\usepackage{maybemath}
\usepackage[hidelinks]{hyperref}

\makeatletter
\newcommand{\mathleft}{\@fleqntrue\@mathmargin0pt}
\newcommand{\mathcenter}{\@fleqnfalse}
\makeatother

\newcommand\p[0]{\partial}

\newcommand{\hf} {\frac{1}{2}}

\newcommand{\nn}{\nonumber\\}

\newcommand{\be}{\begin{equation}}
\newcommand{\ee}{\end{equation}}
\newcommand{\bea}{\begin{eqnarray}}
\newcommand{\eea}{\end{eqnarray}}

\newcommand\fig[1]     {Fig.\,{\ref{#1}}}

\def\eq#1{(\ref{#1})}
\def\s0#1#2{\mbox{\small{$ \frac{#1}{#2} $}}}
\def\0#1#2{\frac{#1}{#2}}

\def\tu{{\tilde u}}

\def\ord#1{{\cal O}(#1)}

\def\mr#1{{\mathrm{#1}}}

\definecolor{garrosgreen}{rgb}{0.1, 0.4, 0.1}
\definecolor{dartmouthgreen}{rgb}{0.05, 0.5, 0.06}
\definecolor{duelferred}{rgb}{0.7, 0.2, 0.1}
\definecolor{cambridgeblue}{rgb}{0.1, 0.3, 1.0}
\definecolor{oxfordblue}{rgb}{0.05, 0.2, 0.7}

\begin{document}
 
\title{Perturbative versus Non--Perturbative Renormalization}

\author{S.~Hariharakrishnan}
\affiliation{University of Debrecen, P.O.~Box 105, H-4010 Debrecen, Hungary}
\affiliation{Universit\"at Hamburg, Bundesstrasse 55, 20146 Hamburg}

\author{U.~D.~Jentschura}
\affiliation{Department of Physics and LAMOR, Missouri University of  Science and Technology, Rolla, Missouri 65409, USA}
\affiliation{HUN-REN Atomki, P.O.~Box 51, H-4001 Debrecen, Hungary}

\author{I.~G.~M\'ari\'an}
\affiliation{HUN-REN Atomki, P.O.~Box 51, H-4001 Debrecen, Hungary}
\affiliation{University of Debrecen, P.O.~Box 105, H-4010 Debrecen, Hungary}

\author{K.~Szab\'o} 
\affiliation{University of Debrecen, P.O.~Box 105, H-4010 Debrecen, Hungary}

\author{I.~N\'{a}ndori}
\affiliation{University of Debrecen, P.O.~Box 105, H-4010 Debrecen, Hungary}
\affiliation{HUN-REN Atomki, P.O.~Box 51, H-4001 Debrecen, Hungary}

\begin{abstract} 
Approximated
functional renormalization group (FRG) equations lead to regulator-dependent 
$\beta$-functions, in analogy to the scheme-dependence of the perturbative 
renormalization group (pRG) approach. A scheme transformation redefines 
the couplings to relate the $\beta$-functions of the FRG method with an 
arbitrary regulator function to the pRG ones obtained in a given scheme. 
Here, we consider a periodic sine-Gordon scalar field theory in $d=2$ 
dimensions and show that the relation of the FRG and pRG approaches is 
intricate. Although both the FRG and the pRG methods are 
known to be sufficient to obtain the critical frequency 
$\beta_c^2 =8\pi$ of the model independently of the choice of the regulator 
and the renormalization scheme, we show that one has to go beyond the 
standard pRG method (e.g., using an auxiliary mass term) or the Coulomb-gas 
representation in order to obtain the $\beta$-function of the wave function 
renormalization. This aspect makes the scheme transformation non-trivial. 
Comparing flow equations of the  two-dimensional sine-Gordon
theory without any scheme-transformation, i.e., redefinition of 
couplings, we find that the auxiliary mass pRG $\beta$-functions of the minimal 
subtraction scheme can be recovered within the FRG approach with the choice 
of the power-law regulator with $b=2$, which therefore constitutes a preferred choice
for the comparison of FRG and pRG flows.
\end{abstract}

\maketitle 

\tableofcontents

\newpage

%=====================================
\section{Introduction}
%===================================== 

{\color{black} The renormalization group (RG) concept has its roots in
theoretical physics, and it plays a pivotal role in understanding the behavior
of physical systems across different scales.  The idea of scale transformations
and scale invariance has been present in physics for decades.} The exact, 
functional renormalization group (FRG) method has been constructed to perform 
renormalization non-perturbatively \cite{kadanoff,wilson,wh,polch,wett,Mo1994}.
{\color{black} The FRG is an implementation of the RG concept, particularly useful 
when dealing with strongly interacting systems in quantum and statistical field theory. 
It bridges the gap between the known microscopic laws (describing fundamental 
particles and interactions) and the complex macroscopic phenomena (such as phase 
transitions and collective behavior) in physical systems. The FRG combines functional 
methods from quantum field theory with the intuitive RG idea proposed by Wilson
\cite{kadanoff,wilson}. Wilson envisioned a way to interpolate smoothly between the 
microscopic and macroscopic descriptions of physical systems. Figuratively, the FRG 
acts as a microscope with adjustable resolution. Unlike perturbative methods that rely 
on small coupling constants, the FRG is non-perturbative. It doesn't require an expansion 
in a small parameter. Instead, it directly explores the system's behavior across scales
\cite{ball,comellas,BeTeWe,polonyi,De2007,Gi2006}.}

The history of the FRG approach runs back over decades starting from the
Wegner-Houghton RG equation \cite{wh} which is based on the Wilson-Kadanoff
blocking \cite{kadanoff,wilson}, through the Polchinski RG equation
\cite{polch} to the modern form of FRG \cite{wett,Mo1994}.  {\color{black} The
central object in FRG is the scale-dependent effective action (also called
average action or flowing action). It depends on a sliding scale (the RG scale)
that continuously decreases as we move from microscopic to macroscopic scales.
The FRG is based on an exact functional flow equation derived by Wetterich and 
Morris in 1993 which describes the dependence of the effective action on the 
RG scale and reads as} \cite{wett,Mo1994},
\begin{equation} 
\label{frg_wetterich}
k \partial_k \Gamma_k [\varphi] = 
\hf \mathrm{Tr}  \left[\frac{k \partial_k R_k }{R_k + \Gamma^{(2)}_k [\varphi]} \right]
=  \hf \int \frac{d^d p}{(2\pi)^d} \,
\frac{k \, \partial_k R_k(p)}{R_k(p) + \Gamma^{(2)}_k[\varphi]}, 
\end{equation} 
where $\Gamma_k [\varphi] = \int \mr{d}^d x \, {\cal L}_k[\varphi]$ is the effective action 
depending on the running scale $k$, $\Gamma^{(2)}_k [\varphi]$ denotes its second 
functional derivative and the trace Tr stands for an integration over all the degrees of 
freedom of the scalar field~$\varphi$. {\color{black} A regulator (infrared cutoff) $R_{k}$, the 
choice of which is arbitrary within certain limits [see Eqs.~(13)---(15) of Ref.~\cite{Gi2006}],
is introduced to decouple slow modes with low momenta while leaving high momentum 
modes unaffected. The regulator ensures that the effective action captures all relevant 
fluctuations.}  Although the explicit form of the $R_{k}$ regulator function is not fixed, the 
solution of the exact FRG equation \eq{frg_wetterich} is independent of the particular 
choice of $R_{k}$ in the low-energy limit $k\to 0$  \cite{litim_pawlowski}. However, once 
approximations have been used, the solution does depend on the regulator function (in 
the IR limit, too) which weakens the predictive power of the method. Because the FRG 
equation cannot be solved without approximations, the regulator-dependence is a serious 
drawback of the method which requires special attention and the search for optimization 
methods is still an active research field, see e.g.,~Ref.~\cite{braun_opt}.

One could argue that the standard perturbative renormalization group (pRG)
procedure is completely free of any problem similar to the regulator-dependence
of the non-perturbative approach. However, this is not true, since perturbative
RG equations are derived with a particular choice of the renormalization
scheme, and thus, $\beta$-functions (at least higher-loop coefficients) are
scheme-dependent. As an example, for the $\beta$ function of
quantum electrodynamics, only the first two leading coefficients (one-loop and
two-loop) are scheme-independent. An explicit derivation of this fact is given
around Eq.~(19.92) 
in Chap.~19 of Ref.~\cite{JeAd2022book}. It is a natural expectation that the
scheme-dependence of pRG is connected with the regulator-dependence of FRG
method \cite{frg_prg,frg_univesality,scheme_transform}. Indeed, the
universality of the FRG method was discussed in~Ref.~\cite{frg_univesality}.  It was
proven in Ref.~\cite{scheme_transform} that the FRG flow equation admits a
perturbative solution; a scheme transformation was given which was used to
obtain the $\beta$ function of the FRG method with a special choice of the
regulator function from the perturbative $\beta$ function obtained in the
modified minimal subtraction ($\overline{{\rm MS}}$) scheme. The $\beta$
functions of the FRG approach are not universal because the FRG method
leads to a mass-dependent scheme which manifests itself through
the nontrivial coupling of mass. In other words, the explicit mass makes the
relation between the FRG and pRG $\beta$ functions nontrivial. If the quantum
field theory has no explicit mass, this scheme transformation is {\color{black}
usually but always simple. For a single real scalar field it is proven to have a 
trivial form, however, it is not necessary the case for multi-component fields
such as the sigma model which is the continuum limit of the 
n-vector model which consists of $N$ real scalar fields coupled by a $\phi^4$ 
interaction that is symmetric under rotations of the $N$ fields. 
The sigma model has important physical realisations and it has been 
investigated by the FRG method, too, as a recent example see e.g., 
Ref.~\cite{WenYinFuHu} where the linear sigma model is coupled to quarks. 
The connection between the FRG method and the perturbative RG with 
$\overline{{\rm MS}}$ scheme has also been studied in \cite{BaPeZa} where 
a regulator function is proposed which reproduces the results of dimensional 
regularization at one and two loops. Thus, the resulting flow equations can be 
seen as non-perturbative extensions of the $\overline{{\rm MS}}$ scheme.} 
In addition, the connection between dimensional regularization and 
Wilsonian RG in regard of the Naturalness/Hierarchy problem is investigated 
in \cite{BrBrCoDa,BrBrCo}.

Let us demonstrate the nontrivial mass dependence of the FRG method 
for a 4-dimensional $\varphi^4$ scalar field theory;
for this simple choice, we have the Euclidean action 
\begin{equation}
\label{ising}
S = \int d^4 x \left[\hf (\partial_\mu\varphi \, \partial_\mu \varphi) + 
\hf m^2 \varphi^2 + \frac{1}{4!} g_4 \varphi^4 \right].
\end{equation}
(The Einstein summation convention is used.) First, we consider its perturbative
renormalization around the Gaussian fixed point ($m^2 = g_4 = 0$) where the pRG
flow equation for the quartic coupling in the minimal subtraction (MS) scheme,
at 1-loop order has been givne in Eq.~(10.52) of \cite{kleinert} or Eq.~(45) of
\cite{yanagisawa} for $\epsilon \equiv 0$) with $k\equiv \mu$.
Its reads as follows,
\begin{align}
\label{g_prg}
\mbox{(1-loop, MS-scheme:)}
\qquad  & k \partial_k g_{4,k} = \frac{3}{16 \pi^2} g_{4,k}^2 + \ord{g_k^3}, \quad
\nn
 \longrightarrow
\quad
& g_{4,k} = g_{4,\Lambda} \,
\left( 1-\frac{3 g_{4,\Lambda}}{16 \pi2} \ln\left( \frac{k}{\Lambda} \right)
\right)^{-1} \,.
\end{align}
Its solution signals the appearance of the Landau pole at very high energies
while in the low energy ($k\to 0$) limit the quartic coupling tends to zero.

The pRG equation \eq{g_prg} can be compared to its non-perturbative counterpart
obtained from Eq.~\eq{frg_wetterich}.  To determine this, the first step is to use
the derivative expansion of the effective action at the lowest order which is
the local potential approximation (LPA),
\begin{equation}
\label{gradient_lpa}
\Gamma_k[\varphi] = \int d^d x \left[V_k(\varphi) + \hf (\partial_\mu\varphi)^2 \right],
\end{equation}
where the Wetterich equation \eq{frg_wetterich} reduces to a differential equation 
for the scale dependent potential $V_k(\varphi)$  
\begin{equation}
\label{lpa_frg}
k \partial_k {V_k(\varphi)} = 
\hf \int \frac{d^d p}{(2\pi)^d} \, 
\frac{k\partial_k R_k}{R_k + p^2 + \partial^2_\varphi V_k} \, .
\end{equation} 
The momentum integral in the RG equation \eq{lpa_frg} can be performed
analytically with the Litim \cite{opt_rg} and with the sharp-cutoff \cite{wh}
regulators which results in the following forms in $d=4$ dimensions,
\begin{eqnarray}
\label{potential_rg_equiation}
& \mbox{(Litim cutoff:)} \qquad 
 R_k( p)= p^2 \left( \frac{k^2}{p^2} -1 \right) 
\Theta\left(1- \frac{p^2}{k^2}\right) %\,,
\\
& \qquad
\longrightarrow
\quad
k\partial_k V_k(\varphi) = 
\frac{1}{32 \pi^2} k^4 \frac{k^2}{k^2 + \partial^2_\varphi V_k(\varphi)} \,,
\\
& \mbox{(Sharp cutoff:)} \qquad 
 R_k( p) = p^2  \frac{1}{\Theta(p^2/k^2-1)} -1 %\,,
\\
& \qquad
\longrightarrow
\quad
k\partial_k V_k(\varphi) = - \frac{1}{16 \pi^2} k^4 
\ln\left[\frac{k^2+\partial^2_\varphi V_k(\varphi)}{k^2}\right] \, .
\end{eqnarray}
As a next step, by substituting the potential $V_k(\varphi) =  \hf m_k^2
\varphi^2 + \frac{1}{4!} g_{4,k} \varphi^4$, one can derive flow equations for
the couplings $m^2_k$ and $g_k$ using the Taylor expansion of both sides of the
above RG equations.  Thus, the non-perturbative flow equations for the quartic
coupling with the Litim and the sharp-cutoff regulators read as [$6/32 = 3/16$]
\begin{eqnarray}
\label{g_frg_litim_phi4}
& \mbox{(Litim cutoff:)} \qquad 
%\nonumber\\
& k \partial_k g_{4,k} = 
\frac{3}{16 \pi^2} \frac{g_{4,k}^2}{(1+m_k^2/k^2)^3} 
\approx \frac{3}{16 \pi^2} g_{4,k}^2, \\
\label{g_frg_sharp_phi4}
& \mbox{(Sharp cutoff:)} \qquad 
%\nonumber\\
& k \partial_k g_{4,k} = 
\frac{3}{16 \pi^2} \frac{g_{4,k}^2}{(1+m_k^2/k^2)^2} \approx 
\frac{3}{16 \pi^2} g_{4,k}^2.
\end{eqnarray}
The difference is in the power of the expression in the denominator.
Let us eliminate the nontrivial mass-dependence of \eq{g_frg_litim_phi4} and
\eq{g_frg_sharp_phi4} by expanding them around the Gaussian fixed point 
($m_k^2 = g_{4} \equiv 0$). In the leading order, both the Litim and the sharp cutoff
regulator can reproduce the perturbative flow equation \eq{g_prg}. In addition,
it is also evident that the leading-order term of the $\beta$-function is found
to be identical for both the Litim and the sharp-cutoff regulators. One can
show that this holds for any choice of the regulator function.
Thus, the leading-order term is regulator-independent. 
In general, the massless FRG flow equation
for the (dimensionful) coupling $g_{2n}$ of the self-iteraction $\varphi^{2n}$
is regulator-independent in $d=2n$ dimensions. In this case,
there is no need for any scheme transformations because the $\beta$ functions
of the FRG and pRG methods are identical \cite{scheme_transform}. 

If one considers the $\varphi^4$ theory for $d<4$ dimensions, the prefactor of
the $g_{4,k}^2$ term in the  RG equations for the dimensionful $\beta$ functions 
becomes regulator-dependent in the FRG. However, the coupling $g_{4,k}$ of $\varphi^4$
carries a dimension for $d<4$ but the RG flow equation should stand for
dimensionless couplings, so a trivial tree-level term appears in the flow
equation when one switches from the dimensionful $g_{4,k}$ to the dimensionless
$\tilde g_{4,k}$ coupling. This trivial term becomes the leading-order term in
the pRG flow equation, and it is naturally regulator-independent and 
is responsible for the
critical behaviour around the Gaussian fixed point. An alternative
choice is to consider the $\varphi^{6}$ self-interaction in $d=3$ where the coupling
$g_6$ is dimensionless and the RG evolution is
regulator-dependent, but still the scheme
transformation is expected to be simple if there is no explicit mass in the
model. A very suitable choice for the study
of the scheme-dependence would be a theory where the coefficients of the $\beta$
function are regulator-dependent for a dimensionless coupling with no explicit
mass. Here, we suggest the periodic scalar theory (i.e., the sine-Gordon model)
which fulfils these requirements, thus, it seems to be a good test model for
scheme-dependence.

So, let us try to modify the scalar model in $d=2$ dimensions
by extending its essential symmetry to a periodic self-interaction. 
Indeed, replacing the mass term and the quartic interactions by a
cosine of the field variable one arrives at the 2-dimensional sine-Gordon (SG)
scalar model defined by the Euclidean action \cite{sg_coleman}
\begin{equation}
\label{sg}
S = \int d^2 x \left[\hf (\partial_\mu\varphi \, \partial_\mu\varphi)
\;  -  \; u\cos(\beta \varphi)\right].
\end{equation}
In addition to the reflection ($Z_2$) symmetry, the action of the SG model
given in Eq.~\eq{sg} remains unchanged under the transformation 
\begin{equation}
\label{pertr}
\varphi(x) \to \varphi(x) + \frac{2\pi}{\beta} \,,
\end{equation}
and thus, it has another discrete symmetry;
namely, it is periodic in the field variable.
Due to this additional symmetry, one expects changes in the phase structure
compared to the $\varphi^4$ model. Indeed, the SG model has two phases in $d=2$
dimensions; it is known to undergo an infinity-order (topological)
Kosterlitz-Thouless-Berezinski (KTB) \cite{ktb1,ktb2} type phase transition
which is controlled by the critical value of the 
frequency $\beta_c^2 = 8\pi$, which
separates the two phases.  Let us note
that both the periodicity and the reflection
symmetry have been broken spontaneously in the broken phase of the model.
The
frequency $\beta^2$ can be seen as the inverse of the wave function
renormalization which is obtained by the rescaling of the field $\varphi \to
\beta \varphi$. Thus, one can derive flow equations for the Fourier amplitude
and the wave function renormalization ($z = 1/\beta^2$). The Fourier amplitude
is dimensionful but the wave function renormalization is dimensionless and its
flow equation is regulator-dependent. This makes the SG model suitable for 
a detailed study of the regulator-dependence.

The non-perturbative RG generates higher harmonics $\cos(n\beta \varphi)$ like
the $\varphi^{2n}$ terms generated in case of the $\varphi^{4}$ theory.
However, it is known that higher harmonics play no role in the phase structure
of the SG model because the SG model can be mapped onto a two-dimensional
Coulomb or vortex-gas, and there, the vortices (charges) with higher vorticity
(multiple charges) are found to be irrelevant, just like the higher harmonics
in the SG theory.  In addition, the perturbative RG treatment is possible for
the SG model, but not in a traditional way, when the potential is expanded in
Taylor series and only  a finite number of terms are kept.  In the case of the
SG model, either one follows the scenario of Refs.~\cite{sg_coleman,rajaraman}
with all terms of the Taylor expansion summed up, or one uses the idea of an
auxiliary mass term presented in Refs.~\cite{amit,balog_hegedus,yanagisawa}.
The two-dimensional SG model is suitable for a detailed study and comparison of
the scheme- and regulator-dependence.

%=====================================
\section{Massless $\maybebm{\varphi^4}$ and 
$\maybebm{\varphi^6}$ models in $\maybebm{2 \leq d \leq 4}$}
%=====================================

\subsection{Orientation}

Before the discussion of the periodic sine-Gordon model, let us compare the
perturbative and non-perturbative RG study of the $\varphi^4$ and $\varphi^6$ 
polynomial theories in $2 \leq d \leq 4$.
In Ref.~\cite{scheme_transform}, an explicit scheme transformation 
(see Eq.~(17) of \cite{scheme_transform}) is given which 
relates the $\varphi^4$ couplings obtained by the FRG method 
(Litim regulator) and the pRG method 
($\overline{{\rm MS}}$ scheme) in $d=4$ dimensions,
\begin{equation}
\label{transform}
g_{\rm FRG} = g_{\overline{{\rm MS}}} + 
{\cal F} g_{\overline{{\rm MS}}}^2 + \ord{g_{\overline{{\rm MS}}}^3},
\end{equation}
where ${\cal F}$ is a function of the explicit mass and, again, the 
$k\equiv \mu$ identification is made. 
This is a perturbative relation obtained by solving the Wetterich equation 
\eq{frg_wetterich} using the loop expansion. 
Differentiating with respect to $k$ and neglecting higher-order terms,
one obtains  in $d=4$ dimensions,
\begin{eqnarray}
\beta_{\rm FRG} 
&= \beta_{\overline{{\rm MS}}} - 2  \frac{m^2_R}{k^2}{\cal F}' 
g_{\overline{{\rm MS}}}^2 +2{\cal F} 
g_{\overline{{\rm MS}}} \beta_{\overline{{\rm MS}}} 
%\\
= \frac{3 g_{\rm FRG}^2 k^6}{16 \pi^2 \left(k^2+m^2\right)^3} 
+ \ord{g_{\overline{{\rm MS}}}^3}
\nn
& \overset{\mbox{$m \to 0$}}{=}  
\frac{3}{16 \pi^2} \, g_{\rm FRG}^2 + \ord{g_{\overline{{\rm MS}}}^3} \,,
\end{eqnarray}
where the $\beta$ function for the Litim regulator in the FRG approach
\eq{g_frg_litim_phi4} is indeed recovered from the 
perturbative $\overline{{\rm MS}}$ scheme. 
Thus, one expects identical results for the pRG and FRG $\beta$
functions in $d=4$ dimensions for vanishing mass. 
In Ref.~\cite{scheme_transform}, it
is argued that a similar relation exists for all regulator types. Let us
discuss the case of $d<4$ dimensions.

%---------
\subsection{Massless $\maybebm{\varphi^4}$ model in 
$\maybebm{2 \leq d \leq 4}$ dimensions}
%---------

It is illustrative to compare the perturbative and non-perturbative RG studies
of the $\varphi^4$ model in $2 \leq d \leq 4$ dimensions. Here we focus on the
flow equation for the quartic coupling $g$ which is dimensionless in $d=4$
dimensions but carries a dimension if $d<4$. The $\beta$ function, i.e., the
pRG flow equation at 1-loop order in the MS scheme in $d=4-\epsilon$ dimensions
is well known [see, e.g., Eq.~(10.52) of Ref.~\cite{kleinert} or Eq.~(45) of
Ref.~\cite{yanagisawa}] and reads as,
\begin{equation}
\label{g_prg_d}
\mbox{(1-loop, MS-scheme:)} %\\
\qquad
k \partial_k \tilde g_{4,k} 
= -\epsilon \tilde g_{4,k} + \frac{3}{16 \pi^2} \tilde g_{4,k}^2 
+ \ord{\tilde g_{4,k}^3}.
\end{equation}
For the details of the $\epsilon$-expansion we refer to
Refs.~\cite{phi4_epsilon} and  \cite{O(N)_epsilon}.
The important difference between Eq.~\eq{g_prg_d} and Eq.~\eq{g_prg} is the
linear term $-\epsilon \tilde g_k$. The quartic coupling $g_{4,k}$ is
dimensionful in $d<4$ but the flow equation must be given for dimensionless
couplings, so, a trivial tree-level term $(d-4) \, \tilde g_{4,k}$ appears in the
pRG equation when one switches from the dimensionful $g_{4,k}$ to the
dimensionless $\tilde g_{4,k}$.  One notes that {\em a priori},
Eq.~\eq{g_prg_d} valid in the limit 
of vanishing $\epsilon$. However, one can use it to extrapolate 
to the cases $\epsilon = 1$ and $\epsilon =2$.
At least, the tree-level term $(d-4) \tilde g_{4,k}$
is correct in any dimension.  The prefactor of $\tilde g_{4,k}^2$ may depend
on the dimension which is not taken into account in Eq.~\eq{g_prg_d} but as we
argued the prefactor of the leading-order, linear term is correct which means
that Eq.~\eq{g_prg_d} correctly reproduces the critical behaviour around the
Gaussian fixed point but non-linear terms become important 
near the Wilson-Fisher fixed point.

Let us now consider the FRG flow equation for $\tilde g_{4,k}$ in $2 \leq d
\leq 4$ dimensions with various choices for the regulator function. We use the
sharp cutoff which gives results equivalent to the Wegner-Houghton FRG
equation, the Litim regulator where the obtained flow equation 
is related to the Polchinski FRG equation via a Legendre
transformation, and the power-law regulator function which could be 
referred to as the Morris-type regulator~\cite{Mo1994}. 
The FRG flow equations read as,
\begin{align}
\label{g_frg_sharp}
& \mbox{(Sharp cutoff:)} \\
& k \partial_k \tilde g_{4,k} =
(d-4) \tilde g_{4,k} + 
\frac{3 \cdot 2^{-d}}{\pi^{d/2} \Gamma(\frac{d}{2})} \tilde g_{4,k}^2
\overset{\mbox{$d=4$}}{=} \frac{3}{16 \pi^2} \tilde g_{4,k}^2, 
\nonumber\\
\label{g_frg_litim}
& \mbox{(Litim cutoff:)} \\
& k \partial_k \tilde g_{4,k} = 
(d-4) \tilde g_{4,k} + 
\frac{3 \cdot 2^{2-d}}{d \cdot \pi^{d/2} \Gamma(\frac{d}{2})} \tilde g_{4,k}^2
\overset{\mbox{$d=4$}}{=} \frac{3}{16 \pi^2} \tilde g_{4,k}^2,  
\nonumber\\
\label{g_frg_power}
& \mbox{(Power-law (with $b=2$):)} \\
& k \partial_k \tilde g_{4,k} = (d-4) \tilde g_{4,k}
- \frac{3}{16} \frac{ 2^{-d} d}{\pi^{\frac{d}{2} - 1}
\Gamma(\frac{d}{2})} \frac{d-4}{\sin(\frac{\pi}{4} d)} \tilde g_{4,k}^2
\nonumber\\
& \qquad \overset{\mbox{$d=4$}}{=}  \frac{3}{16 \pi^2} \tilde g_{4,k}^2 \,.
\nonumber
\end{align}
Here, $R_k(p) = p^2 (k^2/p^2)^b$ is the power-law regulator.

All of the above equations give the same result for $d = 4$ in both orders of
$\tilde g_{4,k}$, and they also agree with Eq.~\eq{g_prg_d}.  Furthermore, the
linear term is the same across all equations for arbitrary $d$, which was
expected, since it comes from its dimensions.  The lack of $d$ dependence in
the second order term of the perturbative solution Eq.~\eq{g_prg_d} prevents it
to be extended and compared to the corresponding term of the FRG results in
other dimensions in any meaningful way. This is, of course, an expected
outcome, since the perturbative expression was derived around $d = 4$ in the
first place.  Note that in Eq.~\eq{g_frg_power} the factor 
$(d-4)/\sin(\pi~d /4)$ has to taken in the appropriate limit 
for $d \rightarrow 4$.

%---------
\subsection{Massless $\maybebm{\varphi^6}$ model in 
$\maybebm{2 \leq d \leq 4}$ dimensions}
%---------

For the $\varphi^6$ model, 
the coupling $g_6$ is dimensionless in $d=3$ dimensions and the leading
order term of its $\beta$-function is expected to be regulator-dependent. Thus,
we extended our potential as $\tilde V_k(\tilde \varphi) = \frac{1}{4!} \tilde
g_{4,k} \tilde \varphi^4 + \frac{1}{6!} \tilde g_{6,k} \tilde \varphi^6$. The
flow equations for $\tilde g_{4,k}$ are the same as in the previous section (of
course, the $g_6$ coupling appears in these flow equations but not at the
leading order). So, the focus is on the $\beta$-functions of the $g_6$
coupling. We use the same regulator functions as for the $\varphi^4$ model.
The result for the sharp cutoff is
\begin{align}
\label{g6_frg_sharp}
\mbox{(Sharp cutoff:)}  \\
k \partial_k \tilde g_{6,k} &= 2(d-3) \tilde g_{6,k} + 
15 \frac{2^{-d} \pi^{-d/2}}{\Gamma(\frac{d}{2})} 
\tilde g_{4,k} (\tilde g_{6,k} - 2 \tilde g_{4,k}^2) \nn
&=
\left\{ 
\begin{array}{cc}
\displaystyle
\frac{15}{4 \pi^2} \tilde g_{4,k} \tilde g_{6,k} - 
\frac{15}{2 \pi^2} \tilde g_{4,k}^3 
& (d=3) \nn [3.1133ex]
\displaystyle
2 {\tilde g}_{6,k} + \frac{15}{16 \pi^2} 
\tilde g_{4,k} \tilde g_{6,k} - 
\frac{15}{8 \pi^2} \tilde g_{4,k}^3
& (d=4) \end{array} \right. \,.
\end{align}
The result for the Litim cutoff is
\begin{align}
\label{g6_frg_litim}
\mbox{(Limit cutoff:)}  \\
k \partial_k \tilde g_{6,k} &= 2(d-3) \tilde g_{6,k} + 
15 \frac{2^{-d}\pi^{-d/2}}{\Gamma(\frac{d}{2})} 
\frac{4}{d} \tilde g_{4,k} (\tilde g_{6,k} - 3 \tilde g_{4,k}^2) 
\nn
&=
\left\{
\begin{array}{cc}
\displaystyle
\frac{5}{\pi^2} \tilde g_{4,k} \tilde g_{6,k} 
- \frac{15}{\pi^2} \tilde g_{4,k}^3 
& (d=3) \nn [3.1133ex]
\displaystyle
2 \tilde g_{6,k} + \frac{15}{16 \pi^2} 
\tilde g_{4,k} \tilde g_{6,k} - 
\frac{45}{16 \pi^2} \tilde g_{4,k}^3 
& (d=4) \end{array} \right. \,.
\end{align}
For the power-law cutoff, one obtains
\begin{align}
\label{g6_frg_power}
\mbox{(Power-law cutoff:)}  \\
k \partial_k \tilde g_{6,k} &= 2(d-3) \tilde g_{6,k} + 
15 \frac{2^{-d} \pi^{-d/2}}{\Gamma(\frac{d}{2})} \,
\frac{\tilde g_{4,k} \pi}{64} \nn
& \quad \times
\left(\tilde g_{6,k} \frac{4(4-d)d}{\sin(\frac{\pi}{4}d)} - \tilde g_{4,k}^2
\frac{(d-6)(d^2-4)}{\cos(\frac{\pi}{4}d)}\right) 
\nn
&= \left\{
\begin{array}{cc}
\displaystyle
\frac{45}{32 \sqrt{2} \pi} \tilde g_{4,k} \tilde g_{6,k} - 
\frac{225}{128 \sqrt{2} \pi} \tilde g_{4,k}^3 
& (d=3) \nn[3.1133ex]
\displaystyle
2 \tilde g_{6,k} + \frac{15}{16 \pi^2} 
\tilde g_{4,k} \tilde g_{6,k} - \frac{45}{128 \pi} \tilde g_{4,k}^3 
& (d=4) \end{array} \right. \,.
\end{align}
We evaluate the special forms of 
the general expression for $d=3$ and $d=4$ for convenience. One can
imediately see, that unlike the $\varphi^4$ case, we do not have 
a regulator
independence in either dimension. However, if we assume that the two
couplings are equally small, then we can say that the $\tilde g_{6,k}$ term is
the leading, the $\tilde g_{4,k} \tilde g_{6,k}$ term is subleading and the
$\tilde g_{4,k}^3$ term is even smaller (sub-subleading). For $d=4$,
we find regulator independence for the first two orders. For $d=3$,
there is no such general behavior, and one can look for scheme-transformations which
relate the FRG $\beta$-functions (with various choices for the regulator) to
the pRG flow equations with various schemes. However, the mass term
is absent, and it was shown in Ref.~\cite{scheme_transform} that the
scheme-transformation is trivial for vanishing mass in $d=4$ dimensions,
thus one expects trivial scheme transformations in $d=3$ dimensions, too. 
In addition, the polynomial model without the explicit mass has no
direct physical realization. So, let us look for a model where one finds no
explicit mass but important physical applications, and consider whether the
scheme transformation between the FRG and pRG schemes are trivial or not.  An
ideal choice is the SG scalar model in $d=2$ dimensions, where
one of its coupling (the frequency which is related to the inverse of the
wave function renormalization) is dimensionless and its $\beta$ function is
regulator dependent \cite{d2_sg}.

%=====================================
\section{Perturbative and non-perturbative RG study of $\maybebm{O(N)}$ models in $\maybebm{2 \leq d \leq 4}$}
%=====================================
%---------
\subsection{Orientation}
%---------
Before we go into the details of the renormalization of the (periodic) SG scalar 
theory it is illustrative to compare the perturbative and non-perturbative RG study 
of (polynomial)  $\maybebm{O(N)}$ models in $\maybebm{2 \leq d \leq 4}$ 
dimensions. Polynomial and periodic models belong to different universality 
classes and their phase transitions are different, too. Let us study the critical 
behavior around the Wilson-Fisher fixed point of $\maybebm{O(N)}$ models 
in the framework of the perturbative and non-perturbative RG approaches.
{\color{black} Before doing this, we would like to draw the attention of the reader 
to a special case where one considers $\maybebm{O(N)}$ scalar field theories 
with nonpolynomial potentials. For example, it was pointed out in 
Refs.~\cite{nonpoly_1} 
and \cite{nonpoly_2} that the linearized RG equations of Wegner and Houghton 
\cite{wh} supported nonpolynomial normal modes of the 
perturbative RG flow in dimensions $d>2$. The resulting theories reduce to the 
sine-Gordon theory at the $d=2$ boundary. This was further connected to the 
exact RG methodology in  \cite{nonpoly_3}. However, the full picture of why 
these modes do not lead to new universality classes for physical RG behavior 
was not laid out for quite some time, until \cite{nonpoly_4}, and the explanation 
is closely tied to the nontrivial connection between the perturbative and 
non-perturbative theory regimes.
}

%---------
\subsection{Perturbative RG approach for $\maybebm{O(N)}$ models}
%---------
The $\maybebm{O(N)}$, i.e., the $N$-vector model \cite{O(N)_epsilon}
is given by the action
\begin{equation}
S = \int d^d x \left[\frac{1}{2} (\partial_\mu \phi)^2  - V(\phi) \right], \hskip 0.5cm
V(\phi) = \frac{1}{2} m^2 \phi^2 + \frac{1}{4!} g \phi^4
\end{equation}
where $\phi = (\varphi_1, \varphi_2, ...,\varphi_N)$ is an $\maybebm{O(N)}$ multiplet.
The perturbative $\beta$ function (for dimensionless couplings) is well-known, 
see e.g., \cite{O(N)_epsilon, perturb_O(N)} where higher-loop corrections were
taken into account. We give the first two orders, 
\begin{equation}
\beta(\tilde g) = - (4-d) \tilde g  + \frac{N+8}{6} \frac{1}{8\pi^2} \tilde g^2 + ...
\end{equation}
which was shown in a general overview Ref.~\cite{yanagisawa} where $\tilde g$ 
is the dimensionless quartic coupling. For $d<4$ there is a non-trivial fixed point, 
i.e., the Wilson-Fisher fixed point  given by 
\begin{equation}
\tilde g_c =  (4-d) \frac{48 \pi^2}{N+8}.
\end{equation}
Let us note that a more accurate value for this fixed point 
is known in the literature where higher-loop corrections were taken into account, 
see e.g., \cite{O(N)_epsilon, perturb_O(N)}. 
For $d=4$ one finds a trivial fixed point $\tilde g=0$ only. In general, the n-th 
order term in $\beta(\tilde g)$ is given by $n! \tilde g^n$, for example
\begin{equation}
\beta(\tilde g) = - (4-d) \tilde g  + \frac{N+8}{6} \tilde g^2 - \frac{9N+42}{36} \tilde g^3 + ...
\end{equation}
where the factor $1/(8\pi^4)$ is included in $\tilde g$. The critical exponents
around the Wilson-Fisher fixed point can be calculated based on the 
$\beta$ function: $\eta$ is given by e.g. Eq.~(70) of Ref.~\cite{yanagisawa},
\begin{equation}
\eta =  \frac{N+2}{2(N+8)^2} (4-d)^2 + ...
\end{equation}
and $\nu$ is given by e.g. Eq.~(92) of Ref.~\cite{yanagisawa}, 
\begin{equation}
 \nu = \hf+ \frac{N+2}{4(N+8)} (4-d) + ...
\end{equation}
In the mean-field approximation one finds $\nu = 1/2$. 
It is interesting to note that, in $d=2$ dimensions, the perturbative 
RG approach does not exclude the existence of a nontrivial
zero of the $\beta$ function close to, but not equal 
to, zero coupling. This zero would otherwise correspond
to the Wilson-Fisher fixed point  in $d=2$ dimension
(see, e.g., \cite{perturb_O(N)}). While the Mermin--Wagner 
theorem excludes long-range order, in the sense of a 
long-range alignment of spins in the $O(N)$ model,
 the existence of other type of long-range 
ordering could be possible due to hidden order parameters
\cite{cugliandolo}. For example in 2d melting, the translational 
order parameter vanishes at all non-zero temperatures, however, 
the system sustains long-range orientational order at finite 
temperatures, thus, the order parameter associated to 
orientational order is not vanishing \cite{cugliandolo}.

%---------
\subsection{Non-perturbative RG approach for $\maybebm{O(N)}$ models}
%---------
The Wetterich FRG equation provides us the possibility to perform the renormalization 
non-perturbatively, although approximations are required to obtain its solution. For 
example, one can use the gradient expansion \eq{gradient_lpa} where the lowest 
order is the LPA. The Wetterich FRG equation \eq{frg_wetterich} at LPA reduces 
to \eq{lpa_frg} which is a partial differential equation for the potential function
and its great advantage is that various functional form of the potential can be 
easily investigated. For example, it is possible to consider a Taylor expansion of 
the effective potential around its minimum. Keeping only the quadratic and quartic 
terms one finds the following dimensionless potential (with dimensionless couplings
and dimensionless field variable)
\begin{equation}
\label{running_min}
V_{k}(\rho)= \frac{\lambda_{k}}{2}(\rho - \rho_{0,k})^2
\end{equation}
where $\rho = \phi^2/2$. It is possible to relate the values of the couplings 
$m^2$ and $g$ with the values of the coupling $\lambda$ and the running 
minimum $\rho_{0}$. These relations give the correct result for the  Wilson-Fisher 
fixed point, but they are not working for the Gaussian fixed point, which is 
$m^2 = g =0$ because no solution of the fixed point equations for $\rho_{0}$ 
and $\lambda$ has a vanishing $\rho_{0}$. By inserting \eq{running_min} into 
the RG equation \eq{lpa_frg} one can obtain flow equations for the couplings 
for general $d$ and $N$, see e.g.,~\cite{BTW_O(N),dupuis_O(N),sanyi_O(N)}
\begin{eqnarray}
k\partial_k \tilde \rho_{0,k} =
(d-2) \tilde \rho_{0,k}+A_{d}\left(1-N-\frac{3  }{(1 + 2 \tilde \rho_{0,k} \tilde \lambda_k)^2}\right),\\
k\partial_k \tilde \lambda_k =
\tilde \lambda_k \left(4-d+2A_{d} \tilde \lambda_k \left(N-1+\frac{9}{(1 + 2 \tilde \rho_{0,k} \tilde \lambda_k)^3}\right)\right),
\end{eqnarray} 
where we used dimensionless quantities ($\tilde \lambda_k$ and $\tilde \rho_{0,k}$)
and $A_d$ is a constant which depends on the dimension. From these  RG 
flow equations the Wilson-Fisher fixed point can be determined. It was shown 
in Ref.~\cite{trunc_rg} that the value of the minimum $\tilde \rho_{0}$ is well defined 
(positive) for every value of $d$ as long as $d>2$ for every $N$. For $d>4$ 
the solution for $\tilde \lambda$ is negative, and, again, this is true for every $N$. 
Figure~6 of Ref.~\cite{trunc_rg} shows that the minimum value $\tilde \rho_{0}$ 
diverges for $d=2$ for every $N$. The Mermin-Wagner-Coleman theorem 
\cite{Mermin66,Hohenberg67,Coleman73} states that a continuous, i.e., the 
$\maybebm{O(N)}$ symmetry for $N \ge 2$ cannot be broken spontaneously 
in two dimensions. Of course, the $N = 1$ case is different because there the 
symmetry is discrete. Thus, the use of the FRG equation with the functional 
form \eq{running_min} is suitable to produce us results in agreement with the 
Mermin-Wagner-Coleman theorem for $N \ge 2$, i.e., no spurious Wilson-Fisher 
fixed points appear in $d=2$ {\color{black} but fails for the case $N=1$ since
it is well-known that the Ising model has two phases in $d=2$ dimensions, so 
one must find  spontaneous symmetry breaking (SSB) for the 
case $N=1$, but the FRG method with the functional form \eq{running_min} 
signals the absence of SSB also for $N=1$ which is not correct \cite{trunc_rg}. }

As a summary, one can conclude that scalar models with $\maybebm{O(N)}$ 
symmetry do not serve as an ideal choice to compare perturbative and 
non-perturbative renormalization: {\em (i)} the scheme transformation between 
$\beta$-functions is non-trivial due to the presence of the mass term, 
{\em (ii)} the
functional forms of the potential used in the perturbative (expansion around 
the Gaussian fixed point) and the non-perturbative (expansion around the 
running minimum) approaches are different.

%=====================================
\section{Perturbative and non-perturbative RG of the SG and CG models}
%=====================================

%=====================================
\subsection{Perturbative RG study of the SG model without auxiliary mass terms}
%=====================================

Let us first present an overview of the standard perturbative study of the 
SG model by using various regularization methods \cite{murayama} without 
the inclusion of any auxiliary mass terms. {\color{black} Let us note that the 
perturbative renormalization of the SG model usually be done in the massive 
case and then one has to consider the massless limit to obtain RG flow equations. 
We refer to this procedure as ``renormalisation with an auxiliary mass term''. In this
subsection, we overview perturbative results where findings were given without
the use of an auxiliary mass term.} The perturbative RG treatment is based 
on the Taylor expansion of the interaction potential which generates the vertices 
of the theory. In the framework of the conventional perturbation theory, when all 
interaction vertices of the Taylor expansion are treated individually, one must 
truncate the expansion and keep only a finite number of terms but then 
periodicity is violated which is the essential symmetry of the SG model. 
If the interaction vertices of the Taylor expansion are not treated individually, 
and one can relate them to each other and be able to sum up all terms to get 
back the cosine function, periodicity is restored. This procedure can be done
if the renormalisation results in an overall multiplicative factor of the couplings
of each vertices which is the case in $d=2$ dimensions when normal-ordering 
is sufficient to remove UV divergences~\cite{sg_coleman,rajaraman,kleinert}. 
This leads to the determination of critical frequency which separates the phases 
of the model. (A different strategy is when one uses an auxiliary mass term 
\cite{amit,balog_hegedus,yanagisawa} where RG flow equations can be 
derived for the wave function renormalization but this is not discussed in this
section.)

The most important conclusion is that one can obtain the critical frequency
$\beta_c^2 = 8\pi$ of the SG model in the framework of the standard perturbative 
approach at one-loop order. This result is independent of the choice of the 
renormalization scheme. Similarly, one finds no regulator-dependence in the
determination of the critical frequency in the framework of the non-perturbative 
FRG approach \cite{d2_sg}.  It is also important to note that no RG flow equation 
is derived for the frequency, i.e., the wave function renormalization.  One could 
assume that this can be done at higher loop order. However, it can be shown 
that it is not possible in the standard perturbative approach \cite{serone}.
In order to be able to derive RG flow equation for the frequency, one
has to go beyond the standard perturbative treatment. One possible way is to
consider the so-called Coulomb gas representation of the SG model. The other
scenario is to extend the original SG model on the basis of an auxiliary mass 
term, perform its perturbative RG study and then take the massless limit 
of the derived RG flow equation for the frequency. Before we present
an overview of these non-standard perturbative results, let us discuss the 
non-perturbative RG study of the SG model.

%=====================================
\subsection{Non-perturbative RG flow equations of the SG model taken at leading order}
%=====================================

Let us discuss the non-perturbative RG study of the two-dimensional SG model
where the exact RG flow equations are expanded in terms of the Fourier
amplitude and only the leading order (LO) terms are kept.  In order to look for
the solution of the FRG equation \eq{frg_wetterich} beyond LPA, we consider the
following ansatz for the SG model
\begin{equation}
\label{eaans_dimful_2}
\Gamma_k[\varphi] = \int d^2 x 
\left[\hf z_k (\partial_\mu{\varphi})^2 + V_k({\varphi}) \right],
%\\
\qquad 
V_k(\varphi) = u_k \cos(\beta \varphi)
\end{equation}
where the local potential contains a single Fourier mode and the wave function
renormalization $z_k$ is assumed to be field-independent. This approximation is
denoted by LPA' (watch the prime!). 
On the one hand the FRG method retains the symmetries of the
model, so the length of period and the frequency remains unchanged over the RG
flow. On the other hand, the field can be rescaled as $\theta = \beta\varphi$
and the inverse of the squared frequency appears in front of the kinetic term.
So, it can be seen as an RG scale-dependent wave function renormalization. The
conclusion is that either one considers the Fourier amplitude and the wave
function renormalization $(u_k, z_k)$ or the Fourier amplitude and the
frequency $(u_k, \beta^2_k)$ as RG scale-dependent parameters. Thus, the
critical frequency $\beta_c$ and the critical value for the wave function
renormalization $z_c$ are related to each other inversely, $z_c = 1/\beta_c^2 =
1/(8\pi)$.
 
The FRG study of the SG model beyond LPA has been discussed in
\cite{d2_sg,d2_sg_z,d_dim_sg}.  In \cite{d2_sg_z} the wave function
renormalization has both field-dependent and independent parts but the
field-dependent part plays no important role in the phase structure (similarly
to the higher harmonics of the potential), so, here we focus on the
field-independent wave function renormalization. So let us summarize briefly the
main results of \cite{d_dim_sg} since our new results presented in the next
section are based on that work. FRG flow equations for the Fourier amplitude
and wave function renormalization are derived at LPA' level in 
Ref.~\cite{d_dim_sg} and read as follows,
\bea
\label{exact_u_rescaled}
k\partial_k u_k &=&
\int \frac{dp \, p}{2\pi}
\frac{k\partial_k \hat R_k}{u_k}
\left(\frac{\hat P-\sqrt{\hat P^2-u_k^2}}{\sqrt{\hat P^2-u_k^2}}\right),\\
\label{exact_z_rescaled}
k\partial_k \hat z_k &=& \int \frac{dp \, p}{2\pi}
\frac{k\partial_k \hat R_k}{2}
\biggl[
\frac{-u_k^2 \hat P(\partial_{p^2} \hat P+ p^2\partial_{p^2}^2 \hat P)}
{[\hat P^2-u_k^2]^{5/2}} 
+\frac{u_k^2 p^2 (\partial_{p^2} \hat P)^2(4 \hat P^2+u_k^2)}
{2 \, [\hat P^2- u_k^2]^{7/2}}
\biggr] ,
\eea
where $P = z_k p^2+R_k$.  Since the dimensionful frequency is
scale-independent, it is convenient to merge it with the scale-dependent wave
function renormalization $z_k$, so, we introduced rescaled quantities $\hat z_k
= z_k/\beta^2$, $\hat R_k = R_k/\beta^2$ and $\hat P =  P/\beta^2 = \hat z_k
p^2+ \hat R_k$.

A very general regulator function is the so called CSS (Compactly Supported Smooth) 
one \cite{css} which is defined as
\begin{equation}
\label{css_norm}
R_k(p) = \; p^2 r(y), \qquad y=p^2/k^2, 
\nonumber%\\
\end{equation}%
\begin{equation}
r_{\mr{css}}(y) = \;
\frac{\exp[\ln(2) c]-1}{\exp\left[\frac{\ln(2) c y^{b}}
{1 -h y^{b}}\right] -1}  \Theta(1-h y^b) 
%\nonumber\\
= \; \frac{2^c -1}{2^{\frac{c \, y^{b}}{1 -h y^{b}}} -1} \, \Theta(1-h y^b) \,.
\end{equation}
Its advantage is to reproduce all major types of regulators in various limits
of its parameters,
\begin{subequations}
\label{css_norm_limits}
\begin{align}
\label{css_norm_litim}
& \mbox{(Optimised Litim-type:)} 
\nonumber\\
& \qquad \lim_{c\to 0,h\to 1} r_{\mr{css}} = 
\left(\frac{1}{y^b} -1\right) \Theta(1-y^b), \\
\label{css_norm_power}
& \mbox{(Power-law Morris-type:)} 
\nonumber\\
& \qquad \lim_{c\to 0, h \to 0} r_{\mr{css}} =
\frac{1}{y^b}, \\ 
\label{css_norm_exp}
& \mbox{(Exponential Wetterich-type:)} 
\nonumber\\
& \qquad \lim_{c \to 1, h \to 0} r_{\mr{css}}= 
\frac{1}{\exp[\ln(2) y^b]-1} \,.
\end{align}
\end{subequations}
Here, $b$ is a free parameter.  The sharp cutoff-regulator can also be reached
by the CSS (e.g. if one takes the limit $b\to \infty$ for the power-law case)
but this type of regulator cannot be applied beyond LPA because it requires
derivatives of the regulator which is problematic for the sharp cutoff since it
is non-differentiable and not continuous. The Litim-type regulator has similar
problem, although it can be applied at LPA' but not beyond that. The
exponential and the power-law regulators have no such problems, they can be
applied at any order of the gradient expansion. 

Momentum integrals of the FRG equation have to be performed numerically, except
the expanded form of Eqs.~\eq{exact_u_rescaled}, \eq{exact_z_rescaled} around
the Gaussian fixed point where analytical results available.  This requires a
special choice for the regulator function $R_k$ such as the power-law
\cite{Mo1994} regulator, so, in this work we perform calculations by this type
of regulator. In general, the regulator function beyond LPA should be given by
the inclusion (multiplicative approach) or the exclusion (additive approach) of
the field independent wavefunction renormalization $z_k$,
\begin{eqnarray}
\label{regulator_multiplicative_additive}
& \mbox{(multiplicative:)} %\nonumber\\
 \qquad & R_k(p) = z_k \, p^2 \, r(p^2/k^2), \\
& \mbox{(additive:)} %\nonumber\\
 \qquad & R_k(p) = p^2 \, r(p^2/k^2) \,.
\end{eqnarray}

In the multiplicative approach, the rescaled regulator $\hat R_k$ contains the
rescaled wavefunction renormalization $\hat z_k$. In the additive approach, the
frequency can be absorbed by the overall multiplicative constant of the
rescaled regulator or can be chosen arbitrarily since it is a scale-independent
free parameter of the model. It is important to note, that the additive approach
requires the use of the power-law regulator function. The phase structure
should be independent whether we use the multiplicative or additive approaches
and of its parameters such as $b$. 

By using the mass cutoff in the additive approach, i.e. power-law type
regulator with $b=1$,  the momentum integrals of  \eq{exact_u_rescaled} and
\eq{exact_z_rescaled} can be performed and the RG equations reads as
\cite{d2_sg},
\begin{align}
\label{sg_single_b1_exact}
(2+k\partial_k)\tu_k =& \;
\frac{1}{2\pi \hat z_k \tu_k} \left[1 -  \sqrt{1 - \tu_k^2} \right], 
\\
k\partial_k \hat z_k =& \;
-\frac{1}{24\pi} \frac{\tu_k^2}{[1 - \tu_k^2]^\frac{3}{2}} \,,
\end{align}
with the dimensionless coupling $\tu = k^{-2} u$. 

Analytic solutions are always available for the power-law type regulator if one
considers the approximated flow equation where the exact RG equation
\eq{exact_u_rescaled} and \eq{exact_z_rescaled} are expanded in Taylor series
with respect to $u_k$ around zero. In the additive approach, the leading order
flow equations have the following forms 
\begin{align}
& \mbox{(additive, power-law, arbitrary b:)} \hfill \nonumber\\
\label{lin_d2_u}
& \qquad (2+k \partial_k) {\tilde u}_k =
\frac{1}{4 \pi {\hat z}_k} {\tilde u}_k
+ \ord{{\tilde u}^3_k} 
\\
\label{lin_d2_z}
& \qquad k \partial_k {\hat z}_k =
- \frac{c_2(b)}{8\pi}  {\hat z}^{\frac{2}{b}-2}_k {\tilde u}^2_k
+ \ord{{\tilde u}^3_k} \,,
\end{align}
with $c_2(b) = \frac{2\pi(b-2)(b-1)}{3 b^2 \sin[2\pi/b]}$ where $c_2(b) > 0$
for $b>1$. These flow equations result in a KTB type (i.e. infinite order)
phase transition with ${\hat z}_{c} =1/(8\pi)$. In the multiplicative approach
one finds very similar leading order RG flow equations,
\begin{align}
& \mbox{(multiplicative, power-law, arbitrary b:)} \nonumber\\
\label{lin_d2_u_multi}
& \qquad (2+k \partial_k) {\tilde u}_k = 
\frac{1}{4 \pi {\hat z}_k} {\tilde u}_k
+ \ord{{\tilde u}^3_k} \,, \\
\label{lin_d2_z_multi}
& \qquad k \partial_k {\hat z}_k =
- \frac{c_2(b)}{8\pi}  {\hat z}^{-2}_k {\tilde u}^2_k
+ \ord{{\tilde u}^3_k}.
\end{align}
It is important to note that the leading-order flow equation for the Fourier
amplitude $\tilde u_k$ are the same in the multiplicative and additive
approaches, which is responsible for the determination of the critical value
$1/\hat z_c = 8\pi$ which separates the phases of the model. This is true for
any choices of the regulator function, so the critical value $z_c$ is found to 
be scheme-independent.
Thus, the scaling of dimensionless Fourier amplitude $\tilde u_k$ has been
determined by trivial tree-level ($\sim k^{-2}$) and the 
non-trivial ($\sim k^{1/(4\pi\hat z)}$) scalings in the above two schemes, 
which result in a regulator-independent  expression, $\tilde u_k
\sim k^{-2 + 1/(4\pi\hat z)}$, from which one can read off the critical value
$1/\hat z_c = 8\pi$.  Similarly, prefactors of the leading order flow equations
obtained for the wave function renormalization are the same in the
multiplicative and additive approaches. The difference is due to the power of
the wave function renormalization thus it is regulator-dependent.

A similar approach has been done in  \cite{malard} where the RG flow equations
above (45) of \cite{malard} were obtained by using the Wilson-Kadanoff blocking
relation up to leading order terms and reads as
\begin{equation}
\label{malard}
k\partial_k \tilde u_k =
\left(\frac{\beta_k^2}{4\pi} -2\right) \tilde u_k,  \hskip 0.5cm
k\partial_k \beta_k^2 = 
\frac{3 \beta_k^6 \tilde u_k^2}{4 \pi \Lambda^3} \,,
\end{equation}
where the identifications $g \equiv \tilde u$ and $\partial_l \equiv
-k\partial_k$ are used. In order to compare them to the leading order RG
equations \eq{lin_d2_u},\eq{lin_d2_z}  and  \eq{lin_d2_u_multi},
\eq{lin_d2_z_multi} of the SG model one has to use the following identification
$\beta^2 = 1/\hat z$, where the additive case reads as 
\begin{align}
\label{d_lin_u_b}
& \mbox{(additive, power-law, arbitrary b:)} \nonumber\\
& \qquad k\partial_k \tilde u_k = 
\left[\frac{\beta_k^2}{4\pi} -2\right] \tilde u_k  \\
\label{d_lin_z_b}
& \qquad k\partial_k \beta_k^2 =
\beta_k^2 \left[ \frac{c(b)}{8\pi} \, 
\tilde u_k^2 (\beta_k^2)^{\frac{3b-2}{b}} \right],
\end{align}
and for the multiplicative case one finds
\begin{align}
\label{d_lin_u_b_multi}
& \mbox{(multiplicative, power-law, arbitrary b:)} \nonumber\\
& \qquad k\partial_k \tilde u_k = 
\left[\frac{\beta_k^2}{4\pi} -2\right] \tilde u_k \,,
\\
\label{d_lin_z_b_multi}
& \qquad k\partial_k \beta_k^2 = 
\beta_k^2 \left[ \frac{c(b)}{8\pi} \, \tilde u_k^2 \beta_k^6 \right] \,.
\end{align}
For $b=2$,
one finds agreement between the additive power-law flow equations 
Eqs.~\eq{d_lin_u_b} and~\eq{d_lin_z_b} and Eq.~\eq{malard}. However, in
\cite{malard} the same non-perturbative (Wilson-Kadanoff) RG method was used,
so, all what one can conclude is that a preferred choice for the 
regulator functions has been identified.
Thus, in the next section let us discuss the Coulomb gas
representation of the SG model and its perturbative RG study.

%=====================================
\subsection{Perturbative RG study of the CG model in the dilute gas representation}
%=====================================

The mapping between the CG and SG models holds in arbitrary dimension
\cite{samuel} (and it is exact in case of point-like charges). Thus, the RG
study of the SG model can be directly used to map out the phase structure of
the CG model and vice versa. In the framework of the real (or coordinate) space
RG approach one can use the dilute gas approximation for the CG model which is
equivalent to the low fugacity, i.e., small Fourier amplitude limit of the SG
model. In this RG approach the charges (vortices) are considered as rigid discs
with finite diameter, so, this can be seen as sharp cutoff version performed in
the coordinate space which corresponds to a smooth cutoff version in the
momentum space. The real space RG equations of the CG model in arbitrary
dimension is given in \cite{d_cg} which have the following form in $d=2$
dimensions
\begin{equation}
\label{cg_rg}
\frac{dx}{d \ell} = -x^2 y^2 \,, \qquad
\frac{dy}{d \ell} = -y(x-2).
\end{equation}
By using the following identifications, $\partial_l = -k \partial_k$, $y \sim
{\tilde u}_k$ and $x \sim 1/{\tilde T} \sim 1/{\hat z}_k$ where ${\tilde T}$ is
the temperature and $y$ is the fugacity, Eq.~\eq{cg_rg} can be rewritten as
\begin{equation}
\label{cg_uz}
k\partial_k {\hat z}_k = - c_z {\tilde u}^2_k, \qquad
(2 + k\partial_k) {\tilde u}_k =  c_u \frac{1}{{\hat z}_k} {\tilde u}_k,
\end{equation}
where $c_u$, $c_z$ are constants that depend on the actual choice for the exact
relation between the parameters of the CG and SG models. It is clear that the
SG flow equations of the additive case \eq{lin_d2_u} and \eq{lin_d2_z} in the
limit $b\to 1$ have identical functional form to \eq{cg_uz}. Since the critical
value $1/\hat z_c = 8\pi$ is independent of the actual choice of the regulator,
one has to define $c_u =1/(4\pi)$. 
By the rescaling of the Fourier amplitude, the
value of the other constant $c_z$ can be chosen to be identical to
$c(b=1)/(8\pi) = 1/(24\pi)$ which results in exactly identical flow equations
to the additive power-law case with $b=1$. 
Thus, it suggests the choice $b=1$. Indeed, the power-law regulator is a
smooth cutoff in the momentum-space RG and this is found to be identical to the
sharp-cutoff of the real space RG approach as expected.

Another real space RG study of the CG model can be found in \cite{barkhudarov}
using again the dilute-gas approximation.  The RG equations (3.2.8) and (3.2.9)
of \cite{barkhudarov} can be rewritten in $d=2$ dimensions 
\begin{align}
\label{barkhudarov}
k\partial_k \tilde u_k =& \;
\left[ \frac{K_2}{2} \, \beta_k^2 -2\right] \tilde u_k \,,
\\
k\partial_k \beta_k^2 = & \;
\beta_k^2 \left[ \left(\frac{I_1 K_2}{2} +B\right) \, 
\frac{\tilde u_k^2}{4} \beta_k^6 \right] \,,
\end{align}
where we introduced the following identifications $2z \equiv \tilde u$, 
$\alpha \equiv \beta$ and $\partial_l \equiv -k\partial_k$ where $K_2$, 
$I_1$ and $B$ are constants.

Once the RG flow equations of the multiplicative case are rewritten by using
$\beta_k^2 = 1/\hat z_k$, see Eqs.~\eq{d_lin_u_b_multi} and
\eq{d_lin_z_b_multi}, one finds flow equations identical to \eq{barkhudarov} if
either the coefficient $c(b)$ and/or the rescaling of $\tilde u$ are chosen
properly. The additive case, see Eqs.~\eq{d_lin_u_b} and \eq{d_lin_z_b} gives
identical functional form only in the so called sharp cutoff limit, i.e., $b\to
\infty$ but then the coefficient $c(b=\infty)$ cannot be defined unambigously,
since the sharp cutoff confronts to the derivative expansion (it is problematic
beyond LPA). So, the multiplicative power-law gives better result then the
additive one and again one can conclude that the sharp-cutoff real space RG
corresponds to a smooth cutoff regulator of the momentum RG. 

However, the perturbative RG study of the equivalent CG representation of the
SG model is not the ideal choice to find connections between renormalization
schemes and regulators because one compares different models. A better choice
is the direct RG study of the SG model. Thus, let us discuss in the next
section the perturbative RG approach for the SG model by using an auxiliary
mass term.

%=====================================
\section{Perturbative RG study of the SG model with an auxiliary mass term}
%=====================================

\subsection{Orientation}

The perturbative renormalization of the SG model has been 
investigated by using the inclusion of an auxiliary mass term, 
see e.g.,\cite{trunc_rg_sg,yanagisawa,amit,balog_hegedus}. 
{\color{black}
Thus, the perturbative renormalization of the SG model is done 
in the massive case when one adds an exilic mass term $m_0$ to 
the original periodic SG theory and RG flow equations of the 
massless SG model are given by considering the limit $m_0 \to 0$. 

The basic idea is to extend the original periodic SG model by 
an explicit mass term. Following \cite{amit} one can consider 
the Euclidean Lagrangian
\begin{align}
\label{amit_sg_1}
{\cal{L}} = \frac{1}{2} (\p_\mu \phi)^2  + \frac{1}{2} m_0 \phi^2 
+ \frac{\alpha_0}{\beta_0 a^2}[1-\cos(\beta_0\phi)] \,,
\end{align}
where $m_0$ is an explicit mass serves as an IR regulator
and $a$ is the UV cutoff (of dimension length) and this is related 
to the momentum RG scale $a \sim 1/k$. One can choose
two different strategies. In the first case the correlation functions 
can be calculated by using an UV regularised form
\begin{align}
\label{amit_corr}
G_0(x) = \frac{1}{2\pi} K_0(m_0 \sqrt{x^2 + a^2}) 
\end{align}
where $K_0$ is the modified Bessel function. In the second case,
one can consider the renormalization of the massive SG model 
and treat $m_0$ as an IR regulator take the limit $m_0 \to 0$ at 
the UV regularized level before UV renormalization, see e.g., 
Ref.~\cite{balog_hegedus}. In any case, $\beta$-functions of the 
original SG model is given in the massless limit,  i.e., the 
renormalisation is performed with an auxiliary mass term and 
in this section we briefly review these results.
}

%------------
\subsection{Leading order perturbative results}
%------------

In Ref.~\cite{yanagisawa} one finds a standard pRG approach for the SG 
model but using the method of an auxiliary mass term. The original SG 
model ({\color{black} without the auxiliary mass})  has the following 
parametrisation in Ref.~\cite{yanagisawa},
\begin{align}
\label{yanagasiwa_sg}
S[\phi] = \int d^2 x \left[\frac{1}{2t} (\p_\mu \phi)^2 - \frac{\alpha}{t}\cos(\phi) \right]    
\end{align}
with the following pRG flow equations, see Eqs.~(159) and (160) in
\cite{yanagisawa} where $\tilde \alpha_k = \tilde u_k \beta_k^2$ and $t_k =
\beta_k^2$,
\begin{eqnarray}
\label{yanagasiwa_prg}
& k \frac{d\tilde \alpha_k}{dk} = \tilde \alpha_k \left(\frac{t}{4\pi} -2\right), 
\quad
& k \frac{d t_k}{dk} = \tilde \alpha_k^2 \frac{t}{32\pi}
\qquad 
\Longrightarrow \\
& k \frac{d\tilde u_k}{d k} = \tilde u_k \left(\frac{\beta_k^2}{4\pi} -2\right) 
-  \tilde u_k^3 \frac{\beta_k^4}{32\pi},
\qquad
& k \frac{d\beta_k^2}{d k} = \tilde u_k^2 \frac{\beta_k^6}{32\pi}, 
\end{eqnarray}
which are identical (keeping only the leading order terms) to the additive
power-law flow equations Eqs. \eq{d_lin_u_b}, \eq{d_lin_z_b} with $b=2$. Of
course, one can use again a scheme transformation to relate the FRG flow
equations with arbitrary choice of the regulator to the pRG flow equations.
However, in this case the transformation is not trivial since one has to
rescale the Fourier amplitude by the frequency, so, practically one should
consider different models to be able to relate the FRG and pRG $\beta$
functions. 

Here, we rather propose to fix the model and look for a particular
choice of the regulator by which the FRG flow equations reproduce the
functional form of the pRG $\beta$ functions. Based on this logic, we have
found the additive power-law flow equations Eqs.~\eq{d_lin_u_b}
and \eq{d_lin_z_b}
with $b=2$ a suitable choice. In the next subsection, we discuss results at
next-to-leading order.

%-----------
\subsection{Next-to-leading order perturbative results}
%-----------

Let us consider the renormalization of the SG model at next-to-leading order
\cite{amit,balog_hegedus} where the method of the auxiliary mass term has been
used, so, this is again not the standard perturbative approach. The action used
for this  pRG calculation ({\color{black} without the auxiliary mass}) reads as,
\begin{align}
\label{amit_sg}
S[\phi] = \int d^2x \left[\frac{1}{2} (\p_\mu \phi)^2 - 
\frac{\alpha}{a^2 \beta^2}\cos(\beta \phi) \right]    
\end{align}
where $a$ is a length scale introduced solely to make $\alpha$ dimensionless.
Since $a$ is in coordinate space, it is related to momentum as $a \sim 1/k$. By
using the following parametrizations, the perturbative RG flow equations of
\cite{amit,balog_hegedus} reads as
\begin{eqnarray}
\label{param_PRG}
 &x = 2 \left(\frac{\beta^2}{8 \pi} - 1\right), \qquad
 &y = \frac{\alpha}{4} \qquad \Longrightarrow \\
 &k \frac{dy}{dk} = xy + A_1 y^3, \qquad
 &k \frac{dx}{dk} = y^2 + B_1 y^2x,     
\end{eqnarray}
where $A_1 = 5/4$ and $B_1 = -1$ and it was argued that $B_1 + 2A_1 = 3/2$ is a
universal number.  The above flow equations which are at next-to-leading order
can be rewritten as,
\begin{subequations}
\label{amit_I}
\begin{align}
k \frac{d\alpha_k}{dk} =& \; 
\left(\frac{\beta_k^2}{4 \pi} - 2\right) \alpha_k + \frac{5}{64} \alpha_k^3, 
\\
k \frac{d\beta_k^2}{dk} =& \;
 \frac{\pi}{4} \alpha_k^2 - \frac{\pi}{4} \alpha_k^2  \left(\frac{\beta_k^2}{4 \pi} - 2\right)
\end{align}
\end{subequations}
and by using the identification $\alpha_k = \tilde u_k \beta_k^2$ the pRG flow equations are the followings,
\begin{subequations}
\label{amit_II}
\begin{align}
k \frac{d\tilde u_k}{dk} =& \;
\left(\frac{\beta_k^2}{4 \pi} - 2\right) \tilde u_k + 
\frac{3\pi}{4} \tilde u_k^3 \beta_k^2 \left(\frac{3\beta_k^2}{16 \pi}-1\right),
\\
k \frac{d\beta_k^2}{dk} =& \;
\frac{\pi}{4} \tilde u_k^2 \beta_k^4 \left(3 - \frac{\beta_k^2}{4 \pi}\right).
\end{align}
\end{subequations}
It is evident that the leading-order FRG flow equations,
\eq{d_lin_u_b},\eq{d_lin_z_b} and \eq{d_lin_u_b_multi},\eq{d_lin_z_b_multi} are
not suitable to reproduce the next-to-leading term propotional to $\tilde
u^3_k$ in \eq{amit_II}. Thus, one has to go beyond the leading-order RG flow
equations obtained by the FRG approach in order to compare them to \eq{amit_II}
which is one of our goals in this work. At this point it is important to
clarify our motivation for the study and comparison of higher order terms of
the RG flow equations. One can argue that these terms have no importance since
they do not influence the critical scaling behaviour around the Coleman fixed
point $\tilde u_\star = 0$ and $\beta^2_\star = 8\pi^2$. However, in
Ref.~\cite{balog_hegedus} it has been shown that one can construct a quantity
from certain combination of the coefficients of these terms of the RG flow
which could be universal. Thus, it is relevant to study the flow equations with
NLO terms which is the goal of the next section.

The main result of this section is that one can use the RG study of the SG
model to compare regulator (i.e., scheme) dependence of non-perturbative and
perturbative approaches and our conclusion is that the power-law type regulator
(with $b=2$) of the FRG method gives identical functional form to the
perturbative RG flow equations obtained by the application of the minimal
subtraction scheme.

%=====================================
\section{Non-perturbative RG flow equations of the SG model taken at next-to-leading order}
%=====================================

Let us now calculate next-to-leading order terms for the flow equations
\eq{lin_d2_u},\eq{lin_d2_z} and \eq{lin_d2_u_multi},\eq{lin_d2_z_multi}. 
{\color{black} To do this one has to expand Eqs.~\eq{exact_u_rescaled}, 
\eq{exact_z_rescaled} around the Gaussian fixed point taking into account 
higher order terms in $\tilde u_k$. Then momentum integrals can be 
performed analytically by the use of the power-law regulator \cite{Mo1994} 
similarly to the leading-order case. As we have already discussed, the 
regulator function beyond LPA can be given by the inclusion (multiplicative 
approach) or the exclusion (additive approach) of the field independent 
wavefunction renormalization $z_k$. }

In the additive approach, the next to leading order (NLO) FRG flow equations 
have the following forms 
\begin{align}
\label{nlo_d2_u}
& \mbox{(additive, power-law, arbitrary b:)} \nonumber\\
& \qquad (2+k \partial_k) {\tilde u}_k =
\frac{1}{4 \pi {\hat z}_k} {\tilde u}_k
+ c_1(b) {\hat z_k}^{\frac{2}{b}-3} {\tilde u_k}^3 
+ \ord{{\tilde u}^5_k} 
\\
\label{nlo_d2_z}
& \qquad k \partial_k {\hat z}_k =
- \frac{c_2(b)}{8\pi}  {\hat z}^{\frac{2}{b}-2}_k {\tilde u}^2_k
- c_3(b) {\hat z_k}^{\frac{4}{b}-4} {\tilde u_k}^4
+ \ord{{\tilde u}^5_k} \,.
\end{align}
The coefficients are as follows,
$c_1(b)=\frac{ (b-2) (b-1)}{8 b^3 \sin[2 \pi /b]}$,
$c_2(b) = \frac{2\pi(b-2)(b-1)}{3 b^2 \sin[2\pi/b]}$, 
and $c_3(b) = \frac{(b-2) (b-1) (b-4) (3 b-4)}{12 b^4 \sin[4 \pi /b]}$.
It is important to note that $c_i(b) > 0$ for $b>1$. 
In the multiplicative approach one finds very similar NLO RG flow equations,
\begin{align}
\label{nlo_d2_u_multi}
&\mbox{(multiplicative, power-law, arbitrary b:)} 
\nonumber\\
&\qquad (2+k \partial_k) {\tilde u}_k = \frac{1}{4 \pi {\hat z}_k} {\tilde u}_k
%\nonumber \\
%&
+ c_1(b) \left(1-\frac{1}{12\pi{\hat z_k}} \right) {\hat z_k}^{-3} {\tilde u_k}^3 
+ \ord{{\tilde u}^5_k} 
\\
\label{nlo_d2_z_multi}
&\qquad k \partial_k {\hat z}_k = - \frac{c_2(b)}{8\pi}  {\hat z_k}^{-2} {\tilde u}^2_k
%\nonumber \\
%&
- c_3(b) \left(1-\frac{c_4(b)}{\hat z_k}\right) {\hat z_k}^{-4} {\tilde u_k}^4
+ \ord{{\tilde u}^5_k} \,,
\end{align}
with $c_4(b)=\frac{(b-2) (b-1) \cot[{2 \pi }/{b}]}{(b-4) b (3 b-4)}$.

It is useful to rewrite these equations in terms of the frequency $\hat
\beta^2_k=1/{\hat z_k}$.  In the additive case one finds
\begin{align}
\label{nlo_d2_u_beta}
& \mbox{(additive, power-law, arbitrary b:)} \nonumber\\
& \qquad k \partial_k {\tilde u}_k = 
\left(\frac{\hat \beta^2_k}{4 \pi } -2 \right) {\tilde u}_k
+ c_1(b) {\hat \beta_k}^{6-\frac{4}{b}} {\tilde u_k}^3 
+ \ord{{\tilde u}^5_k} 
\\
\label{nlo_d2_z_beta}
& \qquad k \partial_k {\hat \beta^2}_k =
 \frac{c_2(b)}{8\pi}  {\hat \beta}^{8-\frac{4}{b}}_k {\tilde u}^2_k
+ {c_3(b)} {\hat \beta_k}^{12-\frac{8}{b}} {\tilde u_k}^4
+ \ord{{\tilde u}^5_k} \,,
\end{align}
while in the multiplicative case
\begin{align}
\label{nlo_d2_u_multi_beta}
& \mbox{(multiplicative, power-law, arbitrary b:)} \nonumber\\
& \qquad k \partial_k {\tilde u}_k = 
\left(\frac{\hat \beta^2_k}{4 \pi } -2 \right) {\tilde u}_k
+ c_1(b) \left(1-\frac{\hat \beta^2_k}{12\pi} \right) {\hat \beta_k}^{6} {\tilde u_k}^3 
%\nonumber\\
%& \qquad \qquad
+ \ord{{\tilde u}^5_k},
\\
\label{nlo_d2_z_multi_beta}
& \qquad k \partial_k {\hat \beta^2}_k =
\frac{c_2(b)}{8\pi}  {\hat \beta_k}^{8} {\tilde u}^2_k
+ {c_3(b)} \left(1-{c_4(b)}{\hat \beta^2_k}\right) {\hat \beta_k}^{12} {\tilde u_k}^4
%\nonumber\\
%& \qquad \qquad
+ \ord{{\tilde u}^5_k}.
\end{align}
The most important observation is that $\tilde u^3_k$ and $\tilde u^4_k$ terms
are generated by the FRG approach at NLO but these terms do not violate the KTB
type (i.e. infinite order) phase transition at ${\hat z}_{c} =1/(8\pi)$.  The
$\tilde u^3_k$ terms of \eq{nlo_d2_u_beta} and \eq{nlo_d2_u_multi_beta} comes
with certain power of $\beta^2_k$ which differs from \eq{amit_II}. However, the
additive case \eq{nlo_d2_u_beta} with $b=2$ gives identical functional form for
the flow equation of the Fourier amplitude found in \eq{yanagasiwa_prg}.
{\color{black} To demonstrate this, on \fig{fig1} we compare the flow diagram 
based on the pRG equation \eq{yanagasiwa_prg} to that of obtained by the 
FRG equations \eq{nlo_d2_u_beta} and \eq{nlo_d2_z_beta} derived by the 
(additive) power-law regulator with various values for the parameter $b$.
 %
% Figure 1
%
\begin{figure}[ht] 
\begin{center}
\includegraphics[width=0.5\linewidth]{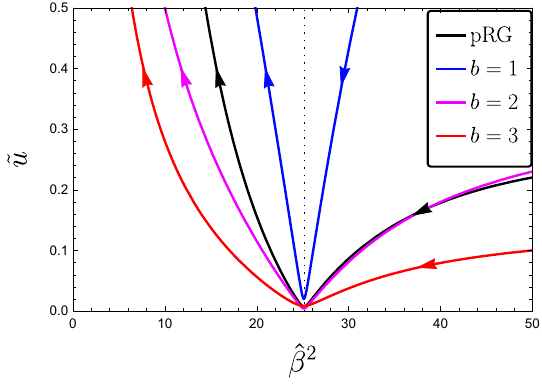}
\caption{RG flow diagram of the SG model indicates the fixed point
$\beta_c^2 = 8\pi$. Black lines are trajectories given by the numerical 
solution of the pRG equation \eq{yanagasiwa_prg}. Colored lines are 
trajectories given by the numerical solution of the FRG equations 
\eq{nlo_d2_u_beta} and \eq{nlo_d2_z_beta} derived by the (additive) 
power-law regulator with various values for the parameter $b$.
\label{fig1}}
\end{center}
\end{figure}
It is clear from \fig{fig1} that RG trajectories of the case $b=2$ are the closest
to the black lines which are the pRG trajectories.}

There are $\tilde u^4_k$ terms in \eq{nlo_d2_z_beta} and \eq{nlo_d2_z_multi_beta}
which are absent in \eq{amit_II}.  Thus, one can conclude that the FRG flow
equations at NLO cannot reproduce the functional form of the perturbative RG
flow equations of \eq{amit_II} obtained at NLO but one finds agreement with
\eq{yanagasiwa_prg} where a $\tilde u^3_k$ term is present in the flow equation
of the Fourier amplitude (although \eq{yanagasiwa_prg} is considered as a
leading order result).

%=====================================
\section{Summary}
%=====================================
In this article, we {\color{black} have attempted} to find connections
between perturbative and non-perturbative renormalization
group equations, with a particular emphasis on the sine--Gordon
model in 2 dimensions.
In \cite{scheme_transform} it {\color{black} was shown} that one can always find a suitable
scheme transformation that maps an FRG $\beta$-function (using a given
regulator) to a perturbative result obtained in a given scheme and vice-versa.
It {\color{black} was argued} in \cite{scheme_transform} that this transformation is {\color{black} 
simplified if there is no explicit mass term in the model (at least for a single field)}. 
It {\color{black} was claimed} that this is due to the fact that the FRG handles the
mass in a non-conventional way, i.e.,~the FRG method 
leads to a mass-dependent scheme which manifests itself through the 
nontrivial coupling of mass.\cite{scheme_transform}. 
Based on these facts, one could argue that
the massless $\phi^{2n}$
model is not a suitable test model for the comparison of perturbative and
non-perturbative renormalization because, {\color{black} of the absence
of the expilict mass Ref.~\cite{scheme_transform}}.

An interesting question is whether one finds this ``trivial'' scheme
transformation for all massless models. With reference to this, {\color{black} here} 
we {\color{black} proposed} to investigate the sine-Gordon scalar model where 
there is no explicit mass term, it is a physically relevant model, and it is known 
that the $\beta$-function obtained for the frequency (the inverse of the wave 
function renormalization) is regulator and scheme-dependent. 
For the sine-Gordon model, concentrating on the 
running of the coupling parameter $u$ alone,
we {\color{black} showed} that in the framework of the standard
perturbative approach, by using dimensional regularization,
one can already obtain the
exact value $\beta_c^2 = 8\pi$ for the critical frequency which separates the
phases of the SG model in $d=2$ dimensions. Furthermore, this result was found
to be independent of the renormalization scheme. This is in perfect agreement
with results of the non-perturbative analysis.  However, we also {\color{black}argued} 
that it is not possible to derive RG flow equation for the wave function
renormalization by the standard perturbative renormalization. Thus, in order to
consider the connection between renormalization schemes and FRG regulators one
has to go beyond the usual, standard perturbative treatment by for example
introducing an auxiliary mass term to the original SG model. In this way, one
can derive RG flow equations for the wave function renormalization, i.e., for
the frequency. For the latter, the $\beta$ function turns out to be scheme 
dependent compare for example Eqs.~\eq{yanagasiwa_prg} 
and \eq{amit_II}.

We {\color{black} have found} that if one relates the FRG and the auxiliary 
mass term pRG $\beta$-functions of the SG model, one obtains 
a non-trivial scheme transformation
which requires the rescaling the Fourier amplitude by the frequency. Although,
it seems to be in contradiction to the general picture based on the results of
\cite{scheme_transform} where any massless model is assumed to require a
``trivial'' scheme transformation it is actually a consequence of the fact that
both the FRG and the auxiliary mass term pRG approaches are mass-dependent
schemes which manifests through the nontrivial coupling of mass.  Thus,
even if the scheme transformation turns out to be somewhat less ``trivial'' 
than initially assumed, strictly speaking, the conclusions of 
Ref.~\cite{scheme_transform} are not violated.

However, if one goes a bit further and compares flow equations obtained by the
FRG and by the auxiliary mass term pRG approaches for the same SG theory,
{\color{black} we showed} that one can choose a particular regulator (power-law
with parameter $b=2$) which gives identical results. In this context, the
power-law (Morris) regulator with $b=2$ turned out to be a choice which leads
to the highest degree of agreement between perturbative and nonperturbative RG
flows. In other words, the power-law regulator with $b = 2$ constituted a
preferred choice for the comparison of FRG and pRG flows, as one compares,
e.g., Eqs.~\eq{yanagasiwa_prg} and \eq{nlo_d2_u_beta}. It is important to note,
that the choice for the regulator {\color{black} suggested} by our analysis in
this work is not a result of an optimisation procedure. However, it is known
(see e.g., Ref.~\cite{jpg}) that the optimal choice for the parameter $b$ of
the power-law regulator is $b=2$.  Indeed, in Ref.~\cite{jpg} the principle of
minimal sensitivity was used for the optimization of the CSS regulator and
confirms that the functional form of a regulator first proposed by Litim is
optimal within the LPA. It {\color{black} was known} that the Litim regulator
provides us critical exponents (for the polynomial scalar field theory)
identical to that of obtained by Polchinski's RG method at LPA.  It was also
known, however, that the Polchinski RG equation has drawbacks in case of the
sine-Gordon \cite{jpg_polchinski} and the multi-layer sine-Gordon models
\cite{aop} at LPA'. In addition, Litim's exact form leads to a kink in the
regulator function, so it confronts with higher order terms of the gradient
expansion. Thus, the choice of the regulator function is an open question
beyond LPA while the results of this work support the use of the power-law
regulator with $b=2$ at LPA'.

%------------------------------------------------------------------
\section*{Acknowledgments}
%------------------------------------------------------------------

We thank Gabor Somogyi for many stimulating discussions on the perturbative 
RG approach for the sine-Gordon theory. U.D.J.~acknowledges support from the 
National Science Foundation (Grant PHY--2110294). The publication was supported 
by the University of Debrecen Program for Scientific Publication.

%\appendix


\vspace*{1cm}

{\bf References}

\vspace*{0.2cm}

\begin{thebibliography}{10}

\bibitem{kadanoff} 
L. P. Kadanoff, Physics {\bf 2}, 263 (1966).

\bibitem{wilson}  
K. G. Wilson, Phys. Rev. D {\bf 3}, 1818 (1971);
K.\ G.\ Wilson and M. \ E.\ Fisher, Phys. Rev. Lett. {\bf 28}, 240 (1972);
K.\ G.\ Wilson, J. Kogut, Phys. Rep. C{\bf 12}, 77 (1974); 
K.\ G.\ Wilson, Rev. Mod. Phys. {\bf  47}, 773 (1975); 
Rev. Mod. Phys. {\bf 55}, 583 (1983). 

\bibitem{wh}
F. J. Wegner, A. Houghton, Phys. Rev. A. {\bf 8}, 401 (1973). 

\bibitem{polch}
J. Polchinski, Nucl. Phys B {\bf 231}, 269 (1984).

\bibitem{wett}
C. Wetterich, Phys. Lett. B {\bf 301}, 90 (1993); 
C. Wetterich, Nucl. Phys. B {\bf 352}, 529 (1991). 

\bibitem{Mo1994}  
T. R. Morris, Int. J. Mod. Phys. A {\bf 9}, 2411 (1994).

%%%%%%%%%%%%%%%%

\bibitem{ball}
R. D. Ball, P. E. Haagensen, J. I. Latorre and E. Moreno,  Phys. Lett. B {\bf 347}, 80 (1995).

\bibitem{comellas}
J. Comellas,  Nucl. Phys. B {\bf 509}, 662 (1998).

\bibitem{BeTeWe}
J. Berges, N. Tetradis, C. Wetterich,  Phys. Rep. {\bf 363}, 223--386 (2002).

\bibitem{polonyi}
J. Polonyi, Central Eur.J.Phys. {\bf 1}, 1-71 (2003).

\bibitem{De2007}
B. Delamotte, 
{\it An Introduction to the Nonperturbative Renormalization Group}, in {\em Springer Notes in Physics vol.~852},
J. Polonyi, A. Schwenk (Eds.), pp.~49--85, Springer (Heidelberg), 2012. See also e-print cond-mat/0702365.

\bibitem{Gi2006}
H. Gies, 
{\it Introduction to the Functional RG and Applications to Gauge 
Theories}, in {\em Springer Notes in Physics vol.~852}, 
J. Polonyi, A. Schwenk (Eds.), pp.~287--348 Springer (Heidelberg), 2012. 
See also e-print hep-ph/0611146.

%%%%%%%%%%%%%%%%

\bibitem{litim_pawlowski}
D. F. Litim and J. M. Pawlowski, Phys. Rev. D {\bf 66}, 025030 (2002).

\bibitem{braun_opt}
N. Zorbach, J. Stoll, and J. Braun, arXiv::2401.12854 [hep-ph].

\bibitem{JeAd2022book}
U.~D. Jentschura and G.~S. Adkins, {\em \relax{Quantum Electrodynamics: Atoms,
  Lasers and Gravity}} (World Scientific, Singapore, 2022).

\bibitem{frg_prg}
A. Codello, J. Phys. A {\bf 45}, 465006 (2012);
U. Ellwanger, Z. Phys. C {\bf 76}, 721 (1997);
S.-B. Liao and J. Polonyi, Ann. Phys. (N.Y.) {\bf 222}, 122 (1993).

\bibitem{frg_univesality}
M. Pernici and M. Raciti, Nucl. Phys. B {\bf 531}, 560 (1998);
A. Bonanno and D. Zappala, Phys. Rev. D {\bf 57}, 7383 (1998);
S. Arnone, A. Gatti, T. R. Morris, and O. J. Rosten, Phys. Rev. D {\bf 69}, 065009 (2004);
O. J. Rosten, Phys. Rep. {\bf 511}, 177 (2012);
T. Papenbrock and C. Wetterich, Z. Phys. C {\bf 65}, 519 (1995).  

\bibitem{scheme_transform}
A. Codello, M. Demmel, and O. Zanusso, Phys. Rev. D {\bf 90}, 027701 (2014).

%%%%%%%%%%%%%%%%

\bibitem{WenYinFuHu}
R. Wen , S. Yin, W. Fu, and M. Huang, Phys. Rev. D {\bf 108}, 076020 (2023).

\bibitem{BaPeZa}
A. Baldazzi, R. Percacci, and L. Zambelli, Phys. Rev. D {\bf 103}, 076012 (2021). 

%%%%%%%%%%%%%%%%

\bibitem{BrBrCoDa}
C. Branchina, V. Branchina, F. Contino, and N. Darvishi, Phys. Rev. D {\bf 106}, 065007 (2022).

\bibitem{BrBrCo}
C. Branchina, V. Branchina, F. Contino, Phys. Rev. D {\bf 107}, 096012 (2023).

%%%%%%%%%%%%%%%%

\bibitem{nonpoly_1}
K. Halpern, K. Huang, Phys. Rev. Lett. {\bf 74}, 3526 (1995).

\bibitem{nonpoly_2}
K. Halpern, K. Huang, Phys. Rev. D 53, 3252 (1996).

\bibitem{nonpoly_3}
V. Periwal, Mod. Phys. Lett. A {\bf 11}, 2915 (1996).

\bibitem{nonpoly_4}
I. H. Bridle and T. R. Morris, Phys. Rev. D {\bf 94}, 065040 (2016).

%%%%%%%%%%%%%%%%

\bibitem{phi4_epsilon}
J. C. Le Guillou and J. Zinn-Justin, Phys. Rev. Lett. {\bf 39}, 95 (1977);
J. C. Le Guillou and J. Zinn-Justin, Phys. Rev. B {\bf 21}, 3976 (1980).

\bibitem{O(N)_epsilon}
R. Guida and J Zinn-Justin, J. Phys. A: Math. Gen. {\bf 31}, 8103 (1998).

\bibitem{kleinert}
H. Kleinert and V. Schulte-Frohlinde, {\em Critical Properties of $\phi^4$-Theories}, World Scientific, https://doi.org/10.1142/4733.

\bibitem{yanagisawa}
T. Yanagisawa, {\em Recent Studies in Perturbation Theory,} Chap.4 (InTech Publisher, 2017) arXiv:1804.02845 [cond-mat.stat-mech];
T. Yanagisawa, Advances in Mathematical Physics, {\bf 2018}, 9238280 (2018). %https://doi.org/10.1155/2018/9238280

\bibitem{opt_rg}  
D. F. Litim, Phys. Lett. B {\bf 486}, 92 (2000); 
{\em ibid}, Phys. Rev. D {\bf 64}, 105007 (2001);  
{\em ibid}, JHEP {\bf 0111}, 059 (2001).  

\bibitem{sg_coleman}
S. Coleman, Physical Review D {\bf 11} 2088 (1975).

\bibitem{ktb1}
J. M. Kosterlitz and D. J. Thouless, J. Phys. C {\bf 6}, 1181 (1973).

\bibitem{ktb2}
V. L. Berezinskii, Zh. Eksp. Teor. Fiz. {\bf 61}, 1144 (1971);
[Sov. Phys. JETP  {\bf 34}, 610 (1972)].

\bibitem{Mermin66}
N. D. Mermin and H. Wagner . Phys. Rev. Lett. {\bf 17}, 1133 (1966).

\bibitem{Hohenberg67}
P. C. Hohenberg, Phys. Rev. {\bf 158}, 383 (1967).

\bibitem{Coleman73}
S. Coleman, Comm. Math. Phys. {\bf 264}, 259 (1973);

\bibitem{perturb_O(N)}
G. A. Baker, B. G. Nickel, and D. I. Meiron, Phys. Rev. B {\bf 17}, 1365 (1978);
A. I. Sokolov, J. Phys. Stud. {\bf 10}, 351 (2006);
M. A. Nikitina and A. I. Sokolov, Phys. Rev. E {\bf 89}, 042146 (2014); 
A. I. Sokolov, M. A. Nikitina, Phys. Rev. E {\bf 89}, 052127 (2014); 

\bibitem{BTW_O(N)}
J. Berges, N. Tetradis, and C. Wetterich, Phys. Rep. {\bf 363}, 223 (2002). 

\bibitem{dupuis_O(N)}
N. Dupuis, Phys. Rev. E  {\bf 83}, 031120 (2011).

\bibitem{sanyi_O(N)}
S. Nagy, Phys. Rev. D {\bf 86}, 085020 (2012).

\bibitem{cugliandolo}
L. F. Cugliandolo,  Lecture note on {\em"Advanced Statistical Physics: Phase Transitions"},
Universite Pierre et Marie Curie, Laboratoire de Physique Theorique et Hautes Energies.
https://www.lpthe.jussieu.fr/~leticia/TEACHING/Master2017/intro-phase-transitions.pdf

\bibitem{trunc_rg}
N. Defenu, P. Mati, I. G. Marian, I. Nandori, A. Trombettoni, JHEP {\bf 05}, 141 (2015).

\bibitem{rajaraman}
R. Rajaraman, Solitons and Instantons (North-Holland, Amsterdam, THe Netherlands, 1982) ISBN: 0-444-87047-4.

\bibitem{murayama}
H. Murayama, {\em Regularization}, "Lawrence Berkeley National Laboratory" (2007), http://hitoshi.berkeley.edu/230A/regularization.pdf.

\bibitem{amit}
D. J Amit, Victor Martin-Mayor, Field Theory, the Renormalization Group, and Critical Phenomena, ISBN: 978-981-256-109-1.

\bibitem{balog_hegedus}
J. Balog, A. Heged\"us, J. Phys. A {\bf 33} 6543 (2000).

\bibitem{d2_sg}
I. Nandori, J. Polonyi, K. Sailer, Phys. Rev. D {\bf 63}, 045022 (2001);
S. Nagy, I. Nandori, J. Polonyi, K. Sailer, Phys. Rev. Letters {\bf 102} 241603 (2009);
I. Nandori, S. Nagy, K. Sailer, A. Trombettoni, Phys. Rev. D {\bf 80}, 025008 (2009);
I. Nandori, S. Nagy, K. Sailer, A. Trombettoni, JHEP {\bf 1009}, 069 (2010);
I. Nandori, I. G. M\'ari\'an, V. Bacs\'o, Phys. Rev.  D {\bf 89}, 047701 (2014). 

\bibitem{serone}
M. Serone, G. Spada and G. Villadoro, %$\lambda \phi^4$ Theory I: The Symmetric Phase Beyond NNNNNNNNLO,
JHEP {\bf 08}, 148 (2018); % [arXiv:1805.05882 [hep-th]]
M. Serone, G. Spada and G. Villadoro, %$\lambda \phi_2^4$ theory \textemdash{} Part II. the broken phase beyond NNNN(NNNN)LO,
JHEP {\bf 05}, 047 (2019). % [arXiv:1901.05023 [hep-th]]

\bibitem{d2_sg_z}
R. Daviet and N. Dupuis, Phys. Rev. Lett. {\bf 122}, 155301 (2019).

\bibitem{d_dim_sg}
I. Nandori, Nucl. Phys. B {\bf 975}, 115681 (2022).

\bibitem{css}
I. Nandori, JHEP {\bf 1304}, 150 (2013).

\bibitem{malard}
M. Malard,  Braz. J. Phys. {\bf 43}, 182 (2013). %arXiv:1202.3481 [cond-mat.str-el]. 

\bibitem{samuel}
S. Samuel, Phys. Rev. D {\bf 18}, 1916 (1978).

\bibitem{d_cg}
J. M. Kosterlitz, J. Phys. C {\bf 10}, 3753 (1977).

\bibitem{barkhudarov}
E. Barkhudarov, Springer Theses, Imperial College London, DOI 10.1007/978-3-319-06154-2 (2014).

\bibitem{trunc_rg_sg}
G. Busiello, L. De Cesare, I. Rabuffo, Phys. Rev. B {\bf 32}, 5918 (1985); 
A. Caramico D'Auria, {\em et al.}, Physica A {\bf 274}, 410 (1999).

\bibitem{jpg}
I. G. M\'ari\'an, U. D. Jentschura, I. N\'andori, J. Phys. G {\bf 41}, 055001 (2014).

\bibitem{jpg_polchinski}
I. N\'andori,  K. Sailer, U. D. Jentschura, G. Soff,  J. Phys. G {\bf 28} (2002) 607. 
      
\bibitem{aop}
U. D. Jentschura, I. N\'andori, J. Zinn-Justin,  Ann. Phys. (N.Y.) {\bf 321}  2647 (2006).
      

\end{thebibliography}
\end{document}